\documentclass[%
 reprint,
 amsmath,amssymb,
 aps,
]{revtex4-1}

\usepackage{graphicx}
\usepackage{dcolumn}
\usepackage{bm}

\usepackage{lineno,hyperref}
\usepackage{epsfig}
\usepackage{float}
\usepackage{bookmark}
\usepackage{times}
\usepackage{amsmath}
\usepackage{color}
\usepackage{epstopdf}
\allowdisplaybreaks[4]
\usepackage{stfloats}
\usepackage{titlesec}
\titleformat{\subsection}[hang]{\normalfont\bfseries}{\thesubsection}{1em}{}[]

\begin{document}

\preprint{APS/123-QED}

\title{When selection pays: structured public goods game with a generalized interaction mode}

\author{Ju Han}

\author{Xiaojie Chen}
\email{xiaojiechen@uestc.edu.cn}
\affiliation{School of Mathematical Sciences, University of Electronic Science and Technology of China, Chengdu 611731, China}

\author{Attila Szolnoki}
\affiliation{Institute of Technical Physics and Materials Science, Centre for Energy Research, P.O. Box 49, H-1525 Budapest, Hungary}

\begin{abstract}
The public goods game is a broadly used paradigm for studying the evolution of cooperation in structured populations. According to the basic assumption, the interaction graph determines the connections of a player where the focal actor forms a common venture with the nearest neighbors. In reality, however, not all of our partners are involved in every games. To elaborate this observation, we propose a model where individuals choose just some selected neighbors from the complete set to form a group for public goods. We explore the potential consequences by using a pair-approximation approach in a weak-selection limit. We theoretically analyze how the number of total neighbors and the actual size of the restricted group influence the critical enhancement factor where cooperation becomes dominant over defection. Furthermore, we systematically compare our model with the traditional setup and show that the critical enhancement factor is lower than in the case when all players are present in the social dilemma. Hence the suggested restricted interaction mode offers a better condition for the evolution of cooperation. Our theoretical findings are supported by numerical calculations.
\end{abstract}

\maketitle

\textbf{To understand the evolution of cooperation in a community of rational individuals is an important scientific challenge. Evolutionary game theory provides a powerful analytical framework for studying this task. More specifically, structured populations offer a realistic approach to model the limited range of communication among participants where a player interacts with nearest neighbors in a public goods game. But what if a player selects just a smaller portion of available partners? We answer this question by introducing a generalized interaction mode where the group size could be arbitrary small in a regular graph. The generalized model offers a better condition for cooperation and the resulting critical enhancement factor is always smaller than the one obtained with the traditional way. We systematically explore how the restricted group size and the full range of nearest neighbors affect the critical enhancement factor separating full defection and full cooperation solutions. When the reduced group size is the smallest and the pair is formed with a neighbor, our theoretical prediction is consistent with the result of prisoner's dilemma game with death-birth strategy update rule.}

\section{Introduction}
The evolution of cooperation is required for the survive and reproduction of many organisms, situations ranging from single-celled bacteria to complex groups of animals, including human society~\cite{Hamilton1963AN,Hamilton1964JTB,Trivers1971JTB,Axelrod1981Science,Smith1995,Hauert2002Science,Fehr2003Nature,Griffin2004Nature,Jusup_PR_2022}. Evolutionary game theory provides an effective theoretical framework for researching how cooperation evolves~\cite{Smith1982,Hofbauer1998,Hofbauer2003BAMS,Nowak2004Nature,Nowak2004Science}. As a typical example of two-person, two-strategy game, the prisoner's dilemma game (PDG) is widely used to model pairwise interactions~\cite{Axelrod1981Science,Nowak1992Nature,Szabo1998PRE,Wahl1999JTB,Santos2005PRL,Duong2021PRSA}. Considering the broad existence of group interactions~\cite{Quan2023Chaos,Liu2023elife}, the public goods games (PGG) have gained popularity as a modeling tool for multi-player game scenarios~\cite{Hauert2002Science,Hauert2006PRSB,Gokhale2010PNAS,Szolnoki2011JTB,Chen2015PRE,Han2015Interface,Quan2019Chaos,Liu2019M3AS,Quan2023MSE}. In particular, when the number of participants are two, the PGG becomes a PDG~\cite{Hauert2002JTB,Hilbe2018Nature}. In both PDG and PGG, rational agents strive to maximize their own benefits, leading to the collapse of cooperation~\cite{Stewart2014PNAS,Sun2023IEEE TNSE}. These two games can work as the paradigms for studying the evolution of cooperation.

Because of the above mentioned advantage of defectors, cooperation can emerge by a specific mechanism. In particular, the network reciprocity is a vital mechanism that enhances cooperative behavior within spatial structure or social network~\cite{Nowak2006Science}. This mechanism has been evidenced effective in both PDG and PGG, supported by a growing body of research~\cite{Lieberman2005Nature,Ohtsuki2006PRSB,Fu2007PA,Chen2008PRE,Santos2008Nature,Van2012JTB,Allen2014EMSS,Zhu2014PA,Perc2013JRSI}. It is well known that cooperation is inhibited in both well-mixed PDG and PGG without additional mechanisms~\cite{Hauert2006PRSB,Hauert2003COM,Ohtsuki2006Nature,Cao2010PHA}. However, Ohtsuki \emph{et al}. studied the cooperative evolution within PDG on a regular graph. They found that under death-birth (DB) update rule, cooperation can emerge when the cost-benefit ratio exceeds a specific critical value~\cite{Ohtsuki2006Nature}. Later Allen \emph{et al}. provided a general formula for weak selection that applies to any population structure~\cite{Allen2017Nature}. Li \emph{et al}. investigated the dynamics of public goods games within structured populations represented by a regular graph. Their findings indicate that population structure can favor the evolution of cooperation and identify the necessary theoretical condition of cooperation under DB update rule~\cite{Li2016PRE}.

We note that the well-mixed population can be depicted by a complete graph. In well-mixed or structured PDG, individuals interact with all the neighbors via the pairwise interaction mode and collect their payoffs. In well-mixed PGG, individuals randomly select some of their neighbors to constitute an interaction group and collect their payoffs~\cite{Hauert2002JTB}. The interaction mode in well-mixed PDG is the same to that in structured PDG. However, in structured PGG, individuals not only participate in interactions with their immediate neighbors, but also engage in indirect interactions with their neighbors' neighbors~\cite{Santos2008Nature}. Indeed, such interaction mode of individuals for structured PGG is different compared with that for well-mixed PGG. When individuals play the PGG in structured populations, they can select some of their neighbors to form an interaction group, similar to the interaction mode in well-mixed PGG. Besides, they may only participate in
these group interactions with the selected neighbors and collect the payoffs due to the limitation of interaction ability. In particular, when such interaction mode is used by individuals in structured PGG, it is identical to that in structured PDG when the participants number is two. Hence it could be a generalized interaction mode. It is unclear under what condition cooperation can emerge in structured PGG when individuals adopt such interaction mode and form a group with just a selected fraction of their potential neighbors.

In this paper, we thus study the evolutionary dynamics of cooperation in structured PGG with such interaction mode. We consider that each individual stochastically chooses some interaction opponents from the neighbors, engage in the group interactions with these selected neighbors, and collect the payoffs. We first derive the dynamical equation for the temporal changes of the frequency of cooperators in the population depicted by a regular network by using the pair approximation approach. Subsequently, we theoretically analyze the stability of the equilibrium points of the dynamical system. We accordingly obtain the theoretical condition in which cooperation can emerge under such interaction mode. We identify the critical value of enhancement factor, above which cooperation can emerge under the interaction mode we considered. We then systematically explore how the critical value of enhancement factor changes with the node degree and the number of group participants. Our analysis shows that this critical value increases as the value of node degree or group number increases. We further demonstrate that our critical value is always smaller than that obtained for the traditional model used in structured PGG~\cite{Li2016PRE}. Moreover, we find that the difference between the critical value we obtained and the one obtained in previous work increase as the node degree increase, but decreases as the group size increases. In addition, we present numerical calculations to verify our theoretical results.

 \section{Model}
We consider the PGG in an infinite large structured population represented by a regular graph with degree $k$. Each player occupies a vertex of the graph and can choose to cooperate or defect. The edges determine who dynamically interacts with whom to gain game payoffs and who competes with whom to achieve biological reproduction. In contrast to well-mixed populations where random interactions happen without spatial constraints, a population structure allows individuals to interact only with their neighbors. In each generation, each individual chooses $n-1$ individuals from its $k$ neighbors (we assume that $n\leq k+1$ in this work). These selected individuals and the focal individual form an $n$-person game group. Therefore, each  individual needs to participate in $\binom{k}{n-1}$ games. Noted that, for $n=2$, such interaction mode in PGG is aligned with the interaction mode  in structured PDG. In the described PGG, each cooperator invests a fixed amount, denoted as cost $c$ ($c>0$), into a common pool. While defectors make no contribution. The total investments made by cooperators are then multiplied by an enhancement factor, $r$ ($r>1$), representing the synergy of their collective efforts. Subsequently, the accumulated amount in the common pool is divided equally among all $n$ group members, independent of their strategies. Accordingly, when there are $i$ cooperators in a group, a cooperator collects the payoff of PGG, which is denoted as
     \begin{equation}
	    \pi_{C}(i)=\frac{i r c}{n}-c,\label{eq1}
     \end{equation}
 and a defector in a group with  $i$ cooperators receives the payoff
     \begin{equation}
	     \pi_{D}(i)=\frac{i r c}{n}.\label{eq2}
     \end{equation}
Furthermore, each individual $i$ participates in $\binom{k}{n-1}$ games and collects the total payoff $P_i$. We set the fitness, $f_i$, of individual $i$, chiefly the reproductive rate, as $1-\omega+\omega P_i$, where  $\omega(0\leq \omega \leq 1)$ measures the strength of selection. In this work, we concentrate on the effects of weak selection~\cite{Fudenberg2006TPB, Nanda2017PNAS}, meaning that $0<\omega \ll 1$. During a microscopic strategy update individuals use the DB update rule, by following previous works~\cite{Ohtsuki2006Nature,Li2016PRE,Ohtsuki2006JTB}. To be specific, an individual is randomly chosen to die at each time step. Subsequently, the vacant site is competed by all its neighbors, who have a probability of acquiring it proportional to their fitness levels. In what follows, we focus on the evolutionary dynamics of cooperation driven by such interaction mode.

\section{Theoretical Analysis and Numerical Results}
Based on the above mentioned description, we can further derive the dynamical equation that describes the dynamical changes of frequency of cooperators in structured populations through the use of pair approximation approach~\cite{Matsuda1992JTB}. Let $x$ denote the frequency of cooperators in the population, we will obtain (for the detailed deviations, see Appendix~\ref{B}).
\begin{equation}
	     \dot{x}=\omega\frac{k-2}{k(k-1)}\binom{k}{n-1}F(k, n, r, c) x(1-x),\label{eq3}
     \end{equation}
where $F(k, n, r, c)=\frac{k+n-1}{n}rc-kc$. Eq.~(\ref{eq3}) indicates that the dynamical system has two equilibrium points, $x = 1$ and $x = 0$ without an internal equilibrium point. We determine that cooperation will be favored by the population if and only if the value of enhancement factor, $r$, meets
     \begin{equation}
	     r>r^*=\frac{k n}{k + n - 1}.\label{eq4}
     \end{equation}
Otherwise, the defectors will eventually occupy the population. It suggests that there exists a critical value of enhancement factor $r^*$, above which cooperation can emerge. In particular, when $n=2$, the PGG become a PDG~\cite{Hilbe2018Nature}. In this case, inequality~(\ref{eq4}) is reduced to
     \begin{equation}
         r>\frac{2k}{k+1}.\label{eq5}
     \end{equation}
Following previous work~\cite{Hilbe2018Nature}, if we use the benefit of cooperation $b$ ($b>0$) and the cost of cooperation $c$ ($c>0$) to characterize the payoff parameters in the PDG, we know that the enhancement factor $r$ for $n=2$ can be represented by
     \begin{equation}
          r=\frac{2b/c}{b/c+1}.\label{eq6}
     \end{equation}
By using Eq.~(\ref{eq6}), inequality~(\ref{eq5}) can further be simplified as
   \begin{equation}
          \frac{b}{c}>k\,,\label{eq7}
      \end{equation}
which is the well-known $b/c>k$ rule~\cite{Ohtsuki2006Nature}. It indicates that, for $n=2$, our theoretical analysis is consistent with that obtained in structured PDG with the DB update rule.

\begin{figure}[h!]
	  \centering
	  \includegraphics[width=3in]{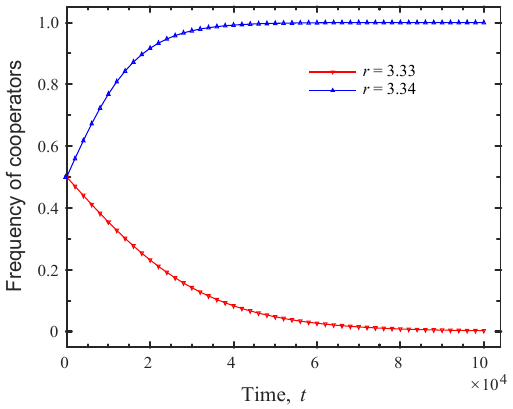}
	   \caption{Frequency of cooperators as a function of time $t$. It is shown that the final state (either
       cooperation or defection) of the population is subject to the value of the enhancement factor $r$. Parameters: $k=8$, $n=5$, $c=1$ and $\omega=0.001$.}
	   \label{fig1}
      \end{figure}

Furthermore, we present numerical calculations to verify the existence of critical enhancement factor that leads to the emergence of cooperation. According to the obtained dynamical equation, we present the frequency of cooperators as a function of time $t$, as shown in Fig.~\ref{fig1}. We here consider a typical regular network---square lattice with Moore's neighborhood~\cite{Szabo2007PR}. We then have $k=8$. We randomly selected 4 individuals from all $8$ neighbors. At this point, the group size of game interactions is denoted as, $n=5$. Our theoretical analysis predicts that the critical value $r^*=\frac{10}{3}$ in this case. We show in Fig.\ref{fig1} for $r=3.34$, the population can eventually evolve to the full cooperation state. On the contrast, for $r=3.33$, cooperation cannot emerge. Instead, defectors take over the whole population. Thus our numerical calculations presented in Fig.~\ref{fig1} support our theoretical predictions.

\begin{figure}
	  \centering
	  \includegraphics[width=3in]{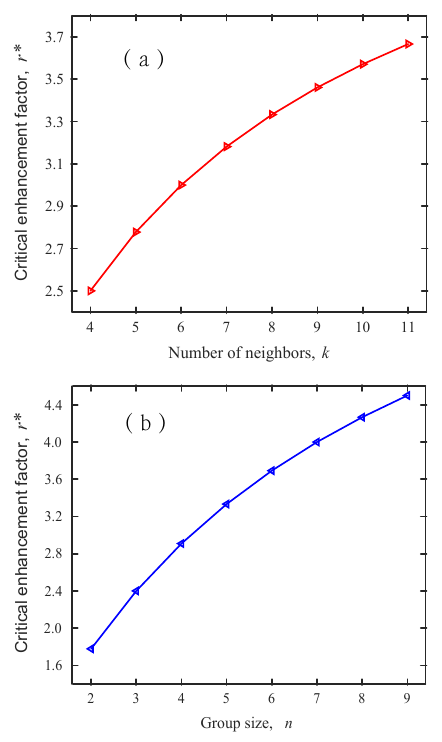}
	  \caption{In panel~(a), the critical enhancement factor, $r^*$, as a function of the number of neighbors $k$ for $n=5$. It is shown that as $k$ increases, there is a monotonically increasing trend in $r^*$. In panel~(b), the critical enhancement factor, $r^*$, as a function of the group size $n$ for $k=8$ . It is also shown that the critical value $r^*$ increases as the value of $n$ increases.}
	\label{fig2}
\end{figure}

\begin{figure}
	\centering
	\includegraphics[width=3in]{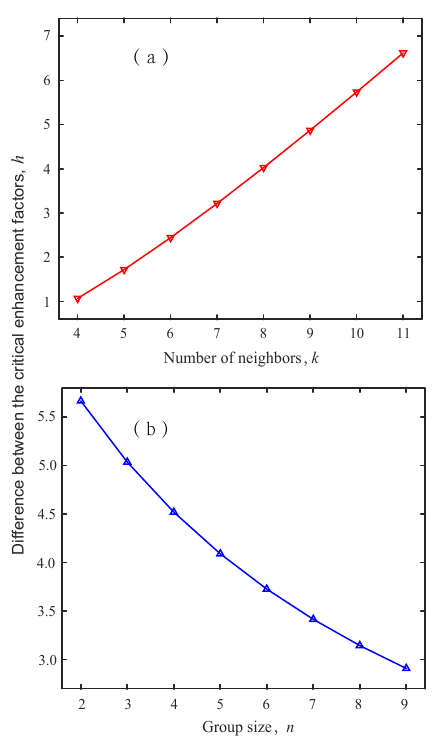}
	\caption{In panel~(a), the difference between the two critical enhancement factors $h$ as a function of $k$ for $n = 5$. It is shown that as $k$ increases, $h$ will also increase. In panel~(b), the difference $h$ as a function of $n$ for $k = 8$. It is shown that as $n$ increases, $h$ will decrease.}
	\label{fig3}
 \end{figure}

Next, we focus on how the critical value of enhancement factor $r^*$ changes with $k$ and $n$, respectively. To do that, we assume that $r^*$ is a binary continuous function of $k$ and $n$. By solving the partial derivative of $r^*$ about $k$, we obtain
     \begin{equation}
          \frac{\partial r^{*}}{\partial k}=\frac{n(n-1)}{(k+n-1)^2}.\label{eq8}
     \end{equation}
Obviously, for $n \ge2$, it is always positive. It indicates that the critical enhancement factor will increase with the increase of $k$. Accordingly, it becomes harder for cooperation to evolve when the neighborhood size $k$ is larger. Furthermore, we validate this finding by numerical calculations performed for $n=5$, as shown in Fig.~\ref{fig2}(a). Simultaneously, by computing the partial derivative of $r^*$ with $n$, we can obtain
     \begin{equation}
          \frac{\partial r^{*}}{\partial n}=\frac{k(k-1)}{(k+n-1)^2}.\label{eq9}
     \end{equation}
When $k\ge2$, we find that the result is also always non-negative. Therefore, we can conclude that as $n$ increases, the $r^*$ value will show a monotonic increasing trend. As a result, cooperation will be much more difficult to be favored when $n$ is larger. Our finding is also confirmed by Fig.~\ref{fig2}(b), where the critical value $r^*$ increases as the group size $n$ increases when the node degree of regular network is set to $k=8$.

We note that if we adopt the previous interaction mode in structured PGG with the DB update rule, there also exists a critical value of enhancement factor, above which cooperation can emerge. Following previous work~\cite{Li2016PRE}, we know that this critical value $r_0^*$ is given as
     \begin{equation}
         r_0^*=\frac{(k+1)^2}{k+3}.\label{eq10}
     \end{equation}
Subsequently, we can compare these two enhancement factors, $r_0^*$ and $r^*$. We accordingly obtain the difference between these two critical values as
     \begin{equation}
	     r_{0}^{*}-r^{*}=\frac{k^3+k^2-(n+1)k+(n-1)}{(k+n-1)(k+3)}.\label{eq11}
     \end{equation}
We can determine that the difference in Eq.~(\ref{eq11}) is always larger than zero for $k\geq n-1$ and $n\geq 2$ (see details in Appendix~\ref{C}). It indicates that cooperation is more likely to emerge when the interaction mode we proposed for structured PGG is used.

In addition, we study how the difference changes  with increasing the two important parameters $k$ and $n$. Let $h(k,n)=r_{0}^{*} -r^{*}$, we can determine that for $k\geq n-1$ the following condition holds
     \begin{equation}
         \frac{\partial h}{\partial k}=\frac{ \alpha }{ \beta }>0,\label{eq12}
     \end{equation}
where
     \begin{align*}
       \alpha  &=k^{4}+(2 n+4) k^{3}+(11 n-6) k^{2} \\
               & +4(n-1) k-\left(4 n^{2}+n-5\right)
     \end{align*}
and
      \begin{equation*}
      \beta=(k+3)^{2}(k+n-1)^{2}.
      \end{equation*}
Accordingly, we can obtain that as $k$ increases, the difference between the critical enhancement factors, $h(k,n)$, will also increase. That is to say, the interaction mode we considered have more advantages for the evolution of cooperation than previous one when the neighborhood size is maximal and all neighbors are involved.

Furthermore, we present the numerical result to verify this theoretical analysis. Let $n=5$, according to Eq.~(\ref{eq11}) and Eq.~(\ref{eq12}), we have
     \begin{equation}
	    h(k,n)=\frac{k^3+k^2-6k+4}{(k+3)(k+4)}\label{eq13}
     \end{equation}
and
     \begin{equation}
	     \frac{\partial h}{\partial k}=\frac{k^{4}+14 k^{3}+49 k^{2}+16 k-100}{(k+3)^{2}(k+4)^{2}}>0\,.\label{eq14}
     \end{equation}
The numerical result, shown in Fig.~\ref{fig3}(a), agrees well with our theoretical prediction. Moreover, we have
     \begin{equation}
	    \frac{\partial h}{\partial n}=-\frac{k(k-1)}{(k+n-1)^{2}}<0,\label{eq15}
      \end{equation}
which indicates that as the value of $k$ increases, the difference between critical enhancement factors $h(k,n)$ will decrease. This finding suggests that as the value of $n$ increases, the effectiveness of the proposed interaction mode decays compared to the previous one. We also present numerical result to verify this theoretical result, as shown in Fig.~\ref{fig3}(b), where we set $k=8$. Accordingly, Eq.~(\ref{eq15}) can be given as
     \begin{equation}
	    \frac{\partial h}{\partial n}=-\frac{56}{(n+7)^{2}}<0.\label{eq16}
     \end{equation}
Our numerical results confirm that albeit the new protocol is still more effective, but the difference decreases as the value of group size $n$ increases.

\section{Discussion and Conclusion}
When exploring the evolutionary dynamics of cooperation among multi-players game, the interaction mode within the structured population adds a considerable level of complexity to the analysis. Meanwhile, it captures a relevant detail of real-life situations and plays an important role in the evolution of cooperation~\cite{Xu2022PRE}. To explore its consequences we here introduce a  generalized interaction mode in structured PGG. We emphasize that the interaction mode we considered is different from the traditional one studied in previous works~\cite{Santos2008Nature,Li2016PRE}. However, this mode is conceptually similar to that used for well-mixed PGG. Meanwhile, it becomes equivalent to the one used for structured PDG when the group size is two. By means of theoretical analysis, we derive the dynamical equation of the frequency of cooperators in the population described by a regular graph. We obtain the mathematical condition in which cooperation can be favored. We determine the critical value of enhancement factor, above which the emergence of cooperation can happen. We emphasize that this critical value is obtained when the accumulative payoff is considered for individuals with the proposed interaction mode. Indeed, it is still the same if the average payoff is used in our work. Moreover, we find that, as the neighborhood size or the group size increases, this critical value will increase and it becomes difficult for the evolution of cooperation. Furthermore, by means of theoretical analysis, we demonstrate that the critical value of enhancement factor we obtain in such interaction mode is always smaller with previous interaction mode in structured PGG. We further obtain that as the node degree increases, the difference of critical value of enhancement factor will increase. However, the difference of the critical value decreases as the group size increases. To complete our study, we have presented numerical results to verify our theoretical analysis.

In this work, we only focus on how cooperation evolves in structured PGG with the interaction mode when the DB  update rule is used. In a future work, it would be interesting to study the evolutionary dynamics when other update rules are considered, such as birth-death (BD) updating, imitation (IM) updating , and pairwise comparison(PC) updating~\cite{Ohtsuki2006JTB}. Furthermore, it is also worthwhile to study the evolutionary dynamics of cooperation when the PGG is played in any population structure with the interaction mode we considered~\cite{Allen2017Nature}.

\begin{acknowledgments}
This research was supported by the National Natural Science Foundation of China (Grant Nos. 62036002 and 61976048) and by the National Research, Development and Innovation Office (NKFIH) under Grant No. K142948.
\end{acknowledgments}

 \section*{Appendix}
 \renewcommand{\thesection}{A}
 \section{The calculation of individual total payoff with different connection patterns.}\label{A}
 \renewcommand{\theequation}{A.\arabic{equation}}
 \setcounter{equation}{0}
In this part, we concentrate on the calculations of individual total payoff with different connection patterns. We first consider the case where the focal individual is a cooperator, who connects with a neighboring defector. To calculate the payoff of cooperator in this case, we assume that there are $k$ nearest neighbors around the cooperator, including $(k-1)q_{C \mid C}$ cooperators and $(k-1)q_{D \mid C}+1$ defectors. Here, $q_{X|Y}$ denotes the conditional probability that a player with strategy $Y$ finds a neighbor with strategy $X$ and $X, Y \in \{C, D\}$. After we randomly choose $i$ individuals from $(k-1)q_{C \mid C}$ cooperators and $n-1-i$ individuals from $(k-1)q_{D \mid C}+1$ defectors. The selected individuals and the focal cooperator form an $n$-person group to play a PGG game. Therefore, according to hypergeometric distribution, the total payoffs of a cooperator can be expressed as
    \begin{equation}
        \pi_{C}^{D} = \sum_{i=0}^{n-1} \binom{(k-1) q_{C|C}}{i} \binom{(k-1) q_{D|C}+1}{n-1-i} \pi_{C}(i+1).\label{eqA1}
    \end{equation}
By utilizing Vandermonde's identity, we can easily obtain
    \begin{equation}
        \sum_{k=0}^{n} \binom{M}{k} \binom{N-M}{n-k}= \binom{N}{n}\label{eqA2}
    \end{equation}
and based on the binomial coefficient relationship, we have
    \begin{equation}
        \binom{N}{n} =\frac{N}{n} \binom{N-1}{n-1}.\label{eqA3}
    \end{equation}
For simplicity, we use notation
    \begin{equation*}
        H(N,M,n,k)=\sum_{k=0}^{n} k\binom{M}{k} \binom{N-M}{n-k}.
    \end{equation*}
Accordingly, we have
    \begin{align}
        \begin{split}
             &H(N,M,n,k)\\
             &=\sum_{k=0}^{n} k \frac{M !}{k !(M-k) !} \frac{(N-M) !}{(n-k) !(N-M-n+k) !} \\
             &=M \sum_{k=1}^{n} \frac{(M-1) !}{(k-1) !(M-k) !} \frac{(N-M) !}{(n-k) !(N-M-n+k) !} \\
             &=M \sum_{j=0}^{n-1} \frac{(M-1) !}{j !(M-1-j) !} \frac{(N-M) !}{(n-1-j) !(N-M-n+j+1) !} \\
             &=M \sum_{j=0}^{n-1} \binom{M-1}{j} \binom{N-M}{n-1-j} \\
             &=M \binom{N-1}{n-1}.\label{eqA4}
        \end{split}
    \end{align}
By substituting  Eq.~(\ref{eqA2}), Eq.~(\ref{eqA3}), and Eq.~(\ref{eqA4}) into Eq.~(\ref{eqA1}), we have
    \begin{align}
        \begin{split}
            \pi_{C}^{D} & =\sum_{i=0}^{n-1}\binom{(k-1) q_{C \mid C}}{i}\binom{k-1) q_{D \mid C}+1}{n-1-i}\pi_{C}(i+1) \\
            & =\sum_{i=0}^{n-1}\binom{(k-1) q_{C \mid C}}{i}\binom{k-1) q_{D \mid C}+1}{n-1-i}
            \left(\frac{(i+1)rc}{n}-c\right) \\
            & =(k-1) q_{C \mid C} \frac{r c}{n} \binom{k-1}{n-2} + (\frac{r c}{n}-c)\binom{k}{n-1}\\
            & =\left((k-1) q_{C \mid C} \frac{r c}{n}+\frac{k}{n-1}(\frac{r c}{n}-c)\right)\binom{k-1}{n-2} \\
            & =\left((k-1) q_{C \mid C} \frac{r c}{n}-\frac{(n-r) k c}{(n-1) n}\right)\binom{k-1}{n-2}.
        \end{split}
    \end{align}
Similarly, we determine a $DD$ pair. We suppose that there are $k$ nearest neighbors around the focal defector, including $(k-1)q_{C \mid D}$ cooperators and $(k-1)q_{D \mid D}+1$ defectors. After we randomly choose $i$ individuals from $(k-1)q_{C \mid D}$ cooperators and $n-1-i$ individuals from $(k-1)q_{D \mid D}+1$ defectors. The selected individuals and the focal defector form an $n$-person group to play the game. Accordingly, the total payoffs of a defector can be written as
    \begin{align}
        \begin{split}
            \pi_{D}^{D} & =\sum_{i=0}^{n-1}\binom{k-1) q_{C \mid D}}{i} \binom{(k-1) q_{D \mid D}+1}{n-1-i} \pi_{D}(i) \\
            & = \sum_{i=0}^{n-1} \binom{k-1) q_{C \mid D}}{i} \binom{(k-1) q_{D \mid D}+1}{n-1-i} \frac{i r c}{n} \\
            & =(k-1) q_{C \mid D} \frac{r c}{n} \binom{k-1}{n-2}.
        \end{split}
    \end{align}

In the third case, we consider a $CC$ pair. Furthermore, we assume that there are $k$ nearest neighbors around a cooperator, including $(k-1)q_{C \mid C}+1$ cooperators and $(k-1)q_{D \mid C}$ defectors. After we randomly choose $i$ individuals from $(k-1)q_{C \mid C}+1$ cooperators and $n-1-i$ individuals from $(k-1)q_{D \mid C}$ defectors. The selected individuals and the focal cooperator form an $n$-person game group.  Therefore, the total payoffs of a cooperator can be expressed as
     \begin{align}
         \begin{split}
             \pi_{C}^{C} & =\sum_{i=0}^{n-1} \binom{(k-1) q_{C \mid C}+1}{i} \binom{(k-1) q_{D \mid C}}{n-1-i} \pi_{C}(i+1) \\
             & =\sum_{i=0}^{n-1}\binom{(k-1) q_{C \mid C}+1}{i} \binom{(k-1) q_{D \mid C}}{n-1-i}  \left(\frac{(i+1)rc}{n}-c\right) \\
             & =\left((k-1) q_{C \mid C}+1\right) \frac{r c}{n} \binom{k-1}{n-2}+(\frac{r c}{n}-c) \binom{k}{n-1}\\
             & =\left(\left((k-1) q_{C \mid C}+1\right) \frac{r c}{n}+\frac{k}{n-1}(\frac{r c}{n}-c)\right)\binom{k-1}{n-2}\\
             & =\left(\left((k-1) q_{C \mid C}+1\right) \frac{r c}{n}-\frac{(n-r)k c}{(n-1) n}\right)\binom{k-1}{n-2}.
         \end{split}
     \end{align}
Last, we determine a $CD$ pair. Moreover, we consider that there are $k$ nearest neighbors around the defector, including $(k-1)q_{C \mid D}+1$ cooperators and $(k-1)q_{D \mid D}$ defectors. After that we randomly choose $i$ individuals from $(k-1)q_{C \mid D}+1$ cooperators and $n-1-i$ individuals from $(k-1)q_{D \mid D}$ defectors. The selected individuals and the focal defector form an $n$-person game group. Therefore, the total payoffs of a defector can be expressed as
     \begin{align}
         \begin{split}
             \pi_{D}^{C} & =\sum_{i=0}^{n-1}\binom{(k-1) q_{C \mid D}+1}{i}\binom{(k-1) q_{D \mid D}}{n-1-i} \pi_{D}(i) \\
             & =\sum_{i=0}^{n-1}\binom{(k-1) q_{C \mid D}+1}{i}\binom{(k-1) q_{D \mid D}}{n-1-i} \frac{irc}{n} \\
             & =\left((k-1) q_{C \mid D}+1\right) \frac{r c}{n}\binom{k-1}{n-2}.
         \end{split}
     \end{align}

 \renewcommand{\thesection}{B}
\section{The derivation process of dynamical equation}\label{B}
\renewcommand{\theequation}{B.\arabic{equation}}
\setcounter{equation}{0}
Based on the calculations of individual total payoff in different cases, we can then derive the dynamical equation of the changes of the frequency of cooperators in regular networks by employing the pair approximation approach~\cite{Ohtsuki2006Nature,Li2016PRE,Sun2023IEEE TNSE}. For this purpose, we employ $p_X$ and $p_Y$ to respectively denote the frequencies of $X$-player and $Y$-player in the population. The frequencies of $XX$, $XY$, $YX$ and $YY$ pairs are represented as $p_{XX}$, $p_{XY}$, $p_{YX}$ and $p_{YY}$. Let $q_{X|Y}$  denote the conditional probability that a player with strategy $Y$ finds a neighbor with strategy $X$. Based on the given hypothesis, when each player has only two strategies to choose from, $X$ and $Y$, we can establish the following equations
     \begin{align}
	     \begin{split}
		     p_{X}+p_{Y}&=1,\\
		     p_{X|X}+p_{Y|X}&=1,\\
		     p_{XY}&=p_{Y}p_{X|Y},\\
		     p_{XY}&=p_{YX}.\label{eqB1}
	     \end{split}
     \end{align}
By utilizing Eqs.~(\ref{eqB1}), only two independent variables remain which are $p_{X}$ and $q_{X|X}$. Below, we deduce the evolutionary dynamical equation of the frequency of cooperators, $p_{C}$, in the whole population.

\renewcommand\thesubsection{(a)}
\subsection{\emph{Updating a defector}}
We here use the death-birth update rule to update a player's strategy. Accordingly, we randomly select a defector to die with probability $p_{D}$. It has $k$ neighbors to compete for the vacant site. Let $i$ and $k-i$ represent the numbers of cooperator and defector among these $k$ neighbors, respectively. Therefore, the frequency of such a configuration is given as
     \begin{equation*}
         \frac{k!}{i!(k-i)!}q_{C \mid D}^{i} q_{D \mid D}^{k-i}.
      \end{equation*}
And the fitness of each cooperator is
     \begin{equation}
         f_{C}=1-\omega+\omega\pi_{C}^{D},
     \end{equation}
where
    \begin{equation*}
            \pi_{C}^{D}=\left((k-1) q_{C \mid C} \frac{r c}{n}-\frac{(n-r) k c}{(n-1) n}\right)\binom{k-1}{n-2}.
    \end{equation*}
Similarly, the fitness of each defector is
     \begin{equation}
         f_{D}=1-\omega+\omega\pi_{D}^{D}.
     \end{equation}
Here
    \begin{equation*}
            \pi_{D}^{D}=(k-1) q_{C \mid D} \frac{r c}{n} \binom{k-1}{n-2}.
    \end{equation*}
The probability that the vacant site is replaced by one of the cooperators is expressed as
     \begin{equation*}
        \frac{if_{C}}{if_{C}+(k-i)f_{D}}.
     \end{equation*}
Under weak selection, expanding the upper equation using Taylor series, we can derive
     \begin{equation}
         \Gamma_C=\frac{i}{k}+\omega\frac{i(k-i)}{k^2}\Pi_{C}+o(\omega^2),
     \end{equation}
where
     \begin{align*}
         \Pi_{C} & = \pi_{C}^{D}-\pi_{D}^{D}\\
                 & = \left((k-1) (q_{C \mid C}-q_{C \mid D}) \frac{r c}{n}-\frac{(n-r) k c}{(n-1)n}\right) \binom{k-1}{n-2}.
     \end{align*}
Therefore, the probability of $p_{C}$ increasing by $1/N$ is written by
     \begin{align}
	     \begin{split}
		     &\operatorname{Prob}\left(\Delta p_{C}=\frac{1}{N}\right)\\
             &=p_{D} \sum_{i=0}^{k}\binom{k}{i} q_{C \mid D}^{i} q_{D \mid D}^{k-i}\frac{if_{C}}{if_{C}+(k-i)f_{D}}.
	     \end{split}
     \end{align}
At the same time, as the individuals embrace the strategy of cooperation, the number of $CC$-pairs will also increase. We know that the number of $CC$-pairs increases by $i$ and the probability of $p_{CC}$ increasing by $i/(kN/2)$  is given as
     \begin{align}
         \begin{split}
		     &\operatorname{Prob}\left(\Delta p_{C C}=\frac{2i}{kN}\right)\\
             &=p_{D}\binom{k}{i} q_{C \mid D}^{i} q_{D \mid D}^{k-i} \frac{if_{C}}{if_{C}+(k-i)f_{D}}.
	     \end{split}
     \end{align}

\renewcommand\thesubsection{(b)}
\subsection{\emph{Updating a cooperator}}
Similarly, we randomly choose a cooperator with the probability of extinction denoted by $p_{C}$. There are $i$ cooperators and $k-i$ defectors in the neighborhood of the empty place. Hence, the frequency of such a configuration is
     \begin{equation*}
         \frac{k!}{i!(k-i)!}q_{C \mid C}^{i} q_{D \mid C}^{k-i}.
     \end{equation*}
The fitness of each cooperator is
     \begin{equation}
         g_{C}=1-\omega+\omega\pi_{C}^{C},
     \end{equation}
where
    \begin{equation*}
             \pi_{C}^{C} =\left(\left((k-1) q_{C \mid C}+1\right) \frac{r c}{n}-\frac{(n-r)k c}{(n-1) n}\right)\binom{k-1}{n-2}.
     \end{equation*}
The fitness of each defector is
    \begin{equation}
        g_{D}=1-\omega+\omega\pi_{D}^{C},
    \end{equation}
where
     \begin{equation*}
             \pi_{D}^{C} =\left((k-1) q_{C \mid D}+1\right) \frac{r c}{n}\binom{k-1}{n-2}.
     \end{equation*}
The probability that the empty space is replaced by one of the defectors is given by
    \begin{equation*}
        \frac{(k-i)g_{D}}{ig_{C}+(k-i)g_{D}}.
    \end{equation*}
Under weak selection, expanding the upper equation using Taylor series, we can easily deduce
     \begin{equation}
         \Gamma_D=\frac{k-i}{k}+\omega\frac{i(k-i)}{k^2}\Pi_{D}+o(\omega^2),
     \end{equation}
where
     \begin{align*}
         \Pi_{D}
         &=\pi_{D}^{C}-\pi_{C}^{C}\\
         & = \left((k-1) (q_{C \mid D}-q_{C \mid C}) \frac{r c}{n}+\frac{(n-r) k c}{(n-1) n}\right)\binom{k-1}{n-2}.
     \end{align*}
Accordingly, the probability of $p_{C}$ decreasing by $1/N$ can be described by
     \begin{align}
	     \begin{split}
		 &\operatorname{Prob}\left(\Delta p_{C}=-\frac{1}{N}\right)\\
         &= p_{C} \sum_{i=0}^{k}\binom{k}{i}q_{C \mid C}^{i} q_{D \mid C}^{k-i}\frac{(k-i)g_{D}}{ig_{C}+(k-i)g_{D}}.
	     \end{split}
     \end{align}
Simultaneously, when the individual takes the defective strategy, the number of $CC$-pairs will decrease. More precisely, the probability of $p_{CC}$ decreasing by $i/(kN/2)$ is denoted as
     \begin{align}
	     \begin{split}
		     &\operatorname{Prob}\left(\Delta p_{C C} = -\frac{2i}{k N}\right)\\
             & = p_{C}\binom{k}{i} q_{C \mid C}^{i} q_{D \mid C}^{k-i} \frac{(k-i)g_{D}}{ig_{C}+(k-i)g_{D}}.
	     \end{split}
     \end{align}

 \renewcommand\thesubsection{(c)}
 \subsection{\emph{Diffusion approximation}}
 We here assume that each evolutionary step takes place in one unit of time, $1/N$. Therefore, the time derivatives of
 $p_{C}$ can be expressed as follows
\begin{align}
   \begin{split}
    \dot{p}_{C} & =\frac{E\left(\Delta p_{C}\right)}{\Delta t} \\
                & =\frac{\frac{1}{N} \operatorname{Prob}\left(\Delta p_{C}=\frac{1}{N}\right)+\left(-\frac{1}{N}\right) \operatorname{Prob}\left(\Delta p_{C}=-\frac{1}{N}\right)}{\frac{1}{N}} \\
                & =p_{D} \sum_{i=0}^{k}\left(\begin{array}{c}
                 k \\
                 i
                 \end{array}\right) q_{C \mid D}^{i} q_{D \mid D}^{k-i} \frac{i f_{C}}{i f_{C}+(k-i) f_{D}} \\
                & -p_{C} \sum_{i=0}^{k}\left(\begin{array}{c}
                k \\
                i
                \end{array}\right) q_{C \mid C}^{i} q_{D \mid C}^{k-i} \frac{(k-i) g_{D}}{i g_{C}+(k-i) g_{D}} \\
                & =p_{D} \sum_{i=0}^{k}\left(\begin{array}{c}
                k \\
                i
                \end{array}\right) q_{C \mid D}^{i} q_{D \mid D}^{k-i} \Gamma_C \\
                & -p_{C} \sum_{i=0}^{k}\left(\begin{array}{c}
                k \\
                i
                \end{array}\right) q_{C \mid C}^{i} q_{D \mid C}^{k-i} \Gamma_D \\
                & =\omega \frac{k-1}{k} p_{C D}\left(q_{D \mid D} \Pi_{C}-q_{C \mid C} \Pi_{D}\right)+o\left(\omega^{2}\right) \\
                & =\omega \frac{k-1}{k}\left(\begin{array}{l}
                k-1 \\
                n-2
                \end{array}\right) G(k, n, r, c)+o\left(\omega^{2}\right) .\label{eqB12}
            \end{split}
          \end{align}
Here,
     \begin{align}
         \begin{split}
	     G(k,n,r,c)
         &=p_{CD}(q_{C \mid C}+q_{D \mid D})\\
         &\cdot\left((k-1) (q_{C \mid C}-q_{C \mid D}) \frac{r c}{n}-\frac{(n-r) k c}{(n-1) n}\right).
         \end{split}
     \end{align}
The time derivatives of $p_{CC}$ can be given by
     \begin{align}
	     \begin{split}
		     \dot{p}_{C C}
             & = \frac{E(\Delta {p}_{CC})}{\Delta t}\\
             & = \left(\sum_{i=0}^{k} \frac{2 i}{k N} \operatorname{Prob}(\Delta p_{C C}=\frac{2 i}{k N}) \right.\\
             & +\left. \sum_{i=0}^{k}(-\frac{2 i}{k N}) \operatorname{Prob}(\Delta p_{C C}=-\frac{2 i}{k N})\right)N\\
             & = p_{D} \sum_{i=0}^{k}\binom{k}{i} q_{C \mid D}^{i} q_{D \mid D}^{k-i}
             \frac{2i}{k}\frac{if_{C}}{if_{C}+(k-i)f_{D}} \\
		     & + p_{C} \sum_{i=0}^{k}\binom{k}{i} q_{C \mid C}^{i} q_{D \mid C}^{k-i}
             (-\frac{2i}{k})\frac{(k-i)g_{D}}{ig_{C}+(k-i)g_{D}} \\
		     & = p_{D} \sum_{i=0}^{k}\binom{k}{i} q_{C \mid D}^{i} q_{D \mid D}^{k-i}
             \frac{2i}{k}(\frac{i}{k}+o(\omega)) \\
		     & + p_{C} \sum_{i=0}^{k}\binom{k}{i} q_{C \mid C}^{i} q_{D \mid C}^{k-i}
             (-\frac{2i}{k})(\frac{k-i}{k}+o(\omega)) \\
             & =p_{D} \sum_{i=0}^{k}\binom{k}{i}\ q_{C \mid D}^{i} q_{D \mid D}^{k-i}  \frac{2i}{k} \frac{i}{k} \\
		     & - p_{C} \sum_{i=0}^{k}\binom{k}{i} q_{C \mid C}^{i} q_{D \mid C}^{k-i}
             \frac{2i}{k}\frac{k-i}{k}+o(\omega) \\
		     & =\frac{2}{k} p_{C D}\left(1+(k-1)\left(q_{C \mid D}-q_{C \mid C}\right)\right)+o(\omega).\label{eqB14}
	     \end{split}
     \end{align}
Furthermore, according to Eq.~(\ref{eqB14}), we can obtain the derivatives of $q_{C \mid C}$ as
     \begin{align}
 	     \begin{split}
 			\dot{q}_{C \mid C} & =\frac{d}{d t}\left(\frac{{p_{C C}}}{p_{C}}\right) \\
            & = \frac{\dot p_{C C}p_{C}-p_{C C}\dot p_{C}}{p_{C}^2}= \frac{\dot p_{C C}}{p_{C}}+o(\omega) \\
 		    & = \frac{2}{k} q_{D \mid C}\left(1+(k-1)\left(q_{C \mid D}-q_{C \mid C}\right)\right)+o(\omega).\label{eqB15}
 	    \end{split}
   \end{align}
As previously mentioned, the system can be represented by $p_C$ and $q_{C|C}$. Rewriting the right-hand of Eq.~(\ref{eqB12}) and Eq.~(\ref{eqB15}) as function of $p_C$ and $q_{C|C}$ can yield the dynamical system
   \begin{align}
     \left\{\begin{array}{l}
         \dot{p}_{C}=w F_{1}\left(p_{C}, q_{C \mid C}\right)+o\left(\omega^{2}\right) . \\
         \dot{q}_{C \mid C}=F_{2}\left(p_{C}, q_{C \mid C}\right)+o(\omega) .
         \end{array}\right.
   \end{align}
Under weak selection, the local frequency of individuals, $q_{C|C}$, reaches equilibrium state much more rapidly compared to the global frequency, $p_C$. Thus the dynamical system quickly converges onto the slow manifold, denoted as $\dot{q}_{C \mid C}=F_{2}(p_{C}, q_{C \mid C}) = 0$. Subsequently, we calculate the quantitative identity of local frequencies at equilibrium and have
     \begin{equation}
          q_{C \mid C}-q_{C \mid D}=\frac{1}{k-1}.
     \end{equation}
Furthermore, by using Eqs.~(\ref{eqB1}), we have the following equations
     \begin{align}
         \begin{split}
		     q_{C \mid C}&=\frac{k-2}{k-1} p_{C}+\frac{1}{k-1}, \\
		     q_{D \mid C}&=\frac{k-2}{k-1}\left(1-p_{C}\right), \\
		     q_{C \mid D}&=\frac{k-2}{k-1} p_{C}, \\
		     q_{D \mid D}&=1-\frac{k-2}{k-1} p_{C},
          \end{split}
     \end{align}
and
     \begin{equation}
	     p_{C D}=\frac{k-2}{k-1} p_{C}\left(1-p_{C}\right).
     \end{equation}
Accordingly, the derivative of $p_C$ with respect to time can be calculated as
     \begin{align}
	     \begin{split}
			\dot{p}_{C} &= \omega\frac{k-1}{k}\binom{k-1}{n-2}G(k, n, r, c)+ o(\omega^2)\\
         &=\omega\frac{k-2}{k(k-1)}\binom{k}{n-1} p_{C}\left(1-p_{C}\right)F(k, n, r, c) + o(\omega^2).
        \end{split}
    \end{align}
Here,
      \begin{equation*}
	     F(k, n, r, c)=\frac{k+n-1}{n}rc-kc.
     \end{equation*}
To simplify some complex forms, we replace $p_C$ by variable $x$ to denote the frequency of cooperators in the population. Thus, the expression can be rephrased as follows
    \begin{equation}
	    \dot{x}=\omega\frac{k-2}{k(k-1)}\binom{k}{n-1} F(k, n, r, c) x(1-x).\label{eqB22}
    \end{equation}
From Eq.~(\ref{eqB22}), we can easily obtain the equilibrium points of the dynamical system, $x=0$ and $x=1$, indicating that the whole population evolves into either full defector or full cooperator state, respectively. To determine the stability of each equilibrium point, we employ the Lyapunov stability analysis method. We can obtain that $x=1$ is only stable if
     \begin{equation*}
         F(k, n, r, c)>0 \Leftrightarrow r>\frac {k n}{k+n-1}.
      \end{equation*}
But if $r<\frac{kn}{k+n-1}$, the equilibrium point $x=0$ is only stable. We thus can make a conclusion that cooperators can occupy the population if and only if
     \begin{equation}
    	r>\frac{k n}{k+n-1}.
     \end{equation}

 \renewcommand{\thesection}{C}
 \section{Comparison of different critical enhancement factors}\label{C}
 \renewcommand{\theequation}{C.\arabic{equation}}
 \setcounter{equation}{0}
 Here we compare the critical enhancement factor $r^{*}$, with the one $r_0^*$ obtained in the traditional interaction model for structured PGG~\cite{Li2016PRE}. We then have
     \begin{align}
         \begin{split}
	     r_{0}^{*}-r^{*}
                         &=\frac{(k+1)^2}{k+3}-\frac{k n}{k+n-1}\\
                         &= \frac{k^3+k^2-(n+1)k+(n-1)}{(k+n-1)(k+3)}\\
                         &=\frac{k\left(k^2-2+k-(n-1)\right)+n-1}{(k+n-1)(k+3)}>0.
         \end{split}
     \end{align}
 Furthermore, we study the difference between $r_{0}^{*}$ and $r^{*}$ as a function of $k$ and $n$. Let $h(k,n)=r_{0}^{*}-r^{*}$ be a binary continuous function. We can then obtain the partial derivative of $h(k,n)$ with respect to $k$, expressed as follows
     \begin{equation}
         \frac{\partial h}{\partial k}=\frac{ \alpha }{ \beta }>0,
     \end{equation}
 where
    \begin{align*}
	\alpha  = & \, k^{4}+(2 n+4) k^{3}+(11 n-6) k^{2} \\& +4(n-1) k-\left(4 n^{2}+n-5\right) \\ \geq & \, k^{4}+(2 n+4)(n-1)+(11 n-6)(n-1) \\& +4(n-1)(n-1)-\left(4 n^{2}+n-5\right) \\ = & \, k^{4}+13 n^{2}-24 n+11>0
    \end{align*}
 and
      \begin{equation*}
      \beta=(k+3)^{2}(k+n-1)^{2}>0.
      \end{equation*}
 Similarly, we have
     \begin{equation}
	    \frac{\partial h}{\partial n}=-\frac{k(k-1)}{(k+n-1)^{2}}<0.
     \end{equation}

\end{document}